\pdfoutput=1
\documentclass[10pt,twocolumn]{article}
\usepackage[a4paper, margin=1.5cm]{geometry}
\usepackage{indentfirst}
\usepackage[utf8]{inputenc}
\usepackage[english]{babel}
\usepackage{amsmath}
\usepackage{amsfonts}
\usepackage{amssymb}
\usepackage[per-mode=symbol]{siunitx}
\usepackage{bm}                       % enhanced bold in math mode
\usepackage{mathpazo}                 % pretty font
\usepackage{microtype}                % better kerning
\usepackage{tikz}
\usepackage{graphicx}
\usepackage{float}                    % to use float placement H
\usepackage[export]{adjustbox}        % to center wide figure
\usepackage{balance}                  % balance columns on last page (\balance)
\usepackage[font=small,labelfont=bf]{caption}
\usepackage{subfig}
\usepackage{ifthen}                   % for if condition in newcommand
\usepackage{booktabs}                 % pretty tables
\usepackage{multirow}                 % fuse cells in tables
\usepackage[symbol]{footmisc}         % symbols for footnotes

\usepackage{hyperref}
\hypersetup{colorlinks=true, linkcolor=blue, citecolor=blue, urlcolor=blue} % NOT SUITED FOR PRINTING!

\usepackage{xcolor}
\definecolor{blue1}{RGB}{ 7,  47,  95}
\definecolor{blue2}{RGB}{18,  97, 160}
\definecolor{blue3}{RGB}{56, 149, 211}
\definecolor{red}{RGB}{210, 0, 0}

\usepackage{titlesec}                 % section customisation
\titlelabel{\thetitle.\ }             % period after section numbers
\titleformat*{\section}{\color{blue1}\scshape\bfseries\centering\large}
\titleformat*{\subsection}{\color{blue2}\normalfont\itshape\large}
\titleformat*{\subsubsection}{\color{blue3}\normalfont\itshape}
\titleformat{\paragraph}[runin]{\color{blue3}\normalfont\itshape}{}{0em}{}[~-]
\titlespacing{\paragraph}{0em}{0em}{0.3em}

\usepackage[backend=bibtex,date=year,doi=false,eprint=false,giveninits=true,isbn=false,maxnames=1,maxbibnames=99,sorting=none,style=nature]{biblatex}
\AtEveryBibitem{
	\clearlist{language}
}
\newlength{\spc} % declare a variable to save spacing value
\let\footnoteorig\footnote
\renewcommand{\footnote}[2]{% #1: footnote text, #2: punctuation
	\ifthenelse{\equal{#2}{,}\OR\equal{#2}{.}}{%
		\settowidth{\spc}{#2}% set value of \spc variable to the width of #2 argument
		\addtolength{\spc}{-1.8\spc}% subtract from \spc about two (1.8) of its values making its magnitude negative
		#2% print the punctuation
		\hspace*{\spc}% print an additional negative spacing stored in \spc after #2
		\footnoteorig{#1}% print the superscript number
	}{%
		\footnoteorig{#1}%
		\ifthenelse{\NOT\equal{#2}{;}\AND\NOT\equal{#2}{:}}{\ }{}%
		#2%
	}%
} % USE sidenoteorig IN SECTIONS!!

\renewcommand{\textcite}[1]{\citeauthor{#1}\hspace*{-0.15em}\supercite{#1}}
\renewcommand{\cite}[2]{% #1: citation string, #2: punctuation
	\ifthenelse{\equal{#2}{,}\OR\equal{#2}{.}}{%
		\settowidth{\spc}{#2}% set value of \spc variable to the width of #2 argument
		\addtolength{\spc}{-1.8\spc}% subtract from \spc about two (1.8) of its values making its magnitude negative
		#2% print the punctuation
		\hspace*{\spc}% print an additional negative spacing stored in \spc after #2
		\supercite{#1}% print (cite) the citation
	}{%
		\supercite{#1}%
		\ifthenelse{\NOT\equal{#2}{;}\AND\NOT\equal{#2}{:}}{\ }{}%
		#2%
	}%
}

\newcommand{\snspace}[2][0.45em]{% #1: punctuation
	\hspace*{-#1}#2\hspace{0.2em}}

\begin{document}

\twocolumn[
	\begin{@twocolumnfalse}

		\begin{center}
			\textbf{\color{blue1}\large Adhesive wear regimes on rough surfaces and interaction of micro-contacts}\\
			\vspace{1em}
			Son Pham-Ba\footnotemark\hspace*{-0.35em},\hspace{0.1em} Jean-François Molinari\\\vspace{0.5em}
			\textit{\footnotesize Institute of Civil Engineering, Institute of Materials Science and Engineering,\\\vspace{-0.2em}
			École polytechnique fédérale de Lausanne (EPFL), CH 1015 Lausanne, Switzerland}
		\end{center}

		% \vspace{0.5em}

		\begin{center}
			% \textit{Abstract}\par
			% \vspace{0.5em}
			\parbox{14cm}{\small
				\setlength\parindent{1em}We develop an analytical model of adhesive wear between two unlubricated rough surfaces, forming micro-contacts under normal load. The model is based on an energy balance and a crack initiation criteria. We apply the model to the problem of self-affine rough surfaces under normal load, which we solve using the boundary element method. We discuss how self-affinity of the surface roughness, and the complex morphology of the micro-contacts that emerge for a given contact pressure, challenge the definition of contact junctions. Indeed, in the context of adhesive wear, we show that elastic interactions between nearby micro-contacts can lead to wear particles whose volumes enclose the convex hull of these micro-contacts. We thereby obtain a wear map describing the instantaneous produced wear volume as a function of material properties, roughness parameters and loading conditions. Three distinct wear regimes can be identified in the wear map. In particular, the model predicts the emergence of a severe wear regime above a critical contact pressure, when interactions between micro-contacts are favored.
				
				\vspace{1em}
			}
		\end{center}

		\vspace{1em}

	\end{@twocolumnfalse}
]

\footnotetext{Corresponding author. E-mail address: \href{mailto:son.phamba@epfl.ch}{son.phamba@epfl.ch}}

\section{Introduction}

Wear is an ubiquitous phenomenon, and yet it is still hardly predictable. It is usual that factors contributing to or limiting wear are studied via extensive  experimental campaigns. Different wear regimes are observed in practice, depending on the ambient and loading conditions, sliding velocity and sliding distance\cite{rabinowiczLeastWear1984,zhangTransitionMildSevere1997}. This paper focuses on the emergence of the severe wear regime in dry contact conditions and assuming adhesive wear processes, for materials of similar hardness.

All macroscopically flat-looking surfaces are in reality rough on a range of scales, whether they are man-made\cite{mandelbrotFractalCharacterFracture1984,majumdarFractalCharacterizationSimulation1990} or natural\cite{thomNanoscaleRoughnessNatural2017}. Often, surface roughness is found to be self-similar or self-affine\cite{perssonContactMechanicsRandomly2006}, in which case it can be described by a fractal dimension or by a Hurst exponent (which describes the scaling of the frequency spectrum\supercite{majumdarFractalCharacterizationSimulation1990}) and the frequency range of self-affinity. These parameters describe the surface geometry and dictate, along with material properties, the contact mechanics\cite{hyunElasticContactRough2007}. When put into contact with each other, two rough surfaces create a number of micro-contacts of various sizes at the interface, depending on the normal load\cite{greenwoodContactNominallyFlat1966,bushElasticContactRough1975,perssonElastoplasticContactRandomly2001,hyunFiniteelementAnalysisContact2004}. The real contact area is, for common loading conditions, a small fraction of the apparent area, and in the small load limit, it was shown to be roughly proportional to the normal load by a wide variety of analytical and numerical models\cite{yastrebovInfinitesimalFullContact2015,muserMeetingContactMechanicsChallenge2017}.

Upon frictional sliding, wear particles are eventually created at these micro-contacts, when enough elastic energy is accumulated at the contact junctions which resist sliding. The most fundamental approach to understanding wear is to first consider a single contact junction. A breakthrough in the understanding of adhesive wear was made thanks to computer simulations that revealed a material length scale at which a transition between ductile and brittle behavior occurs\cite{aghababaeiCriticalLengthScale2016}. This length scale is related to the smallest particle size that can be detached from a contact interface, as first theorized by Rabinowicz\cite{rabinowiczEffectSizeLooseness1958} in 1958. Recent molecular dynamics (MD) simulations revealed that contact junctions above this length scale generate wear particles, while surfaces asperities that form smaller junctions deform plastically\cite{aghababaeiCriticalLengthScale2016}. In a system comprising of two sliding solids with identical material properties, the critical junction diameter for the ductile to brittle transition is defined as 
\begin{equation}\label{eq:d_star_ramin}
	d^* = \Lambda \frac{4\gamma G}{\sigma_\text{j}^2} \,,
\end{equation}
where $\Lambda \sim O(1)$ is a factor depending on the exact geometrical configuration of the system, $G$ is the shear modulus, $\gamma$ is the surface energy, and $\sigma_\text{j}$ is the shear strength of the junction formed between the two colliding asperities, a parameter ultimately linked to the adhesive strength or, in case of full adhesion, to the yield strength. Again, we highlight that a wear particle can only be detached at the junction if $d \geqslant d^*$\snspace. This quantity provides a mechanistic rationale to the transition between low wear (plastic deformation of protruding asperities) and wear particles generation. 

MD simulations also revealed that two nearby contact junctions interact elastically when the distance separating them is of the order of the junction diameter\cite{aghababaeiCriticalLengthScale2016}. Elastic interactions result in a crack shielding mechanism, so that during wear particle formation, not all cracks can properly develop as they are unloaded by nearby propagating cracks, leading to the formation of larger wear particles. In this context, instead of forming two wear particles at the two separate junctions, a single larger wear particle, whose size encloses the two junctions, is created\cite{aghababaeiAsperityLevelOriginsTransition2018}. These elastic interactions were further explored in more details analytically in a two dimensional setting\cite{pham-baAdhesiveWearInteraction2020}. Using an energy balance criterion, a wear map was obtained to predict how a set of multiple tangentially loaded micro-contacts transition from multiple small wear particles to a single larger wear particle. 
Crack shielding mechanisms were also confirmed in the context of finite-element simulations with a phase-field approach to fracture\cite{colletVariationalPhasefieldContinuum2020}. We emphasize that these studies on contact junctions interactions were carried in a rather academic context of 2D or quasi-2D setups. A recent 3D numerical model explored  the effect of elastic interactions for multiple contact junctions, whose shapes were optimized for an energetically efficient wear particle removal process\cite{molinariOptimizedMaterialRemoval2020}.

Elastic interactions between a large number of micro-contacts, with varying sizes and shapes, as occurs during loading of self-affine surfaces, is largely unexplored and is the focus of the present paper. 
Frérot \emph{et al}\cite{frerotMechanisticUnderstandingWear2018}. used the boundary element method (BEM) to numerically simulate the contact between two rough surfaces and obtain a map of the micro-contacts. Each micro-contact size is compared to $d^*$ to assess if it can result in the formation of a wear particle. Ultimately, an instantaneous wear coefficient can be estimated. Brink \emph{et al}\cite{brinkParameterfreeMechanisticModel2020}. followed the same principle but added the notion of sliding distance to compute a wear volume over time and obtain a more physically accurate wear coefficient. These models give promising results, but their downside is that they do not take into account elastic interactions and assume that each micro-contact is isolated from the others, overlooking the potential transition to a severe wear regime. An interesting study by Popov and Pohrt\cite{popovAdhesiveWearParticle2018} also relies on BEM to compute the contact between rough surfaces. They use an energy balance criterion to determine if a wear particle can be detached, therefore accounting for elastic interactions. However, we will discuss how the approximation that was made in the released energy, while allowing for a high computational efficiency, does not capture the full effect of the elastic interactions. As a consequence, they obtain wear particles that can enclose several junctions, but do not observe a transition to a severe wear regime.

In summary, this paper explores a mesoscale mechanistic model for adhesive wear. In the context of self-affine surfaces in contact, the model aims to explain the emergence of different wear regimes as function of normal load. Section~\ref{sec:theory} extends the two-dimensional model of Pham-Ba \emph{et al}\cite{pham-baAdhesiveWearInteraction2020}. to three dimensions. To this end, we describe analytically the elastic interactions between nearby micro-contacts and discuss how these affect the formation of wear particles. Then, in Section~\ref{sec:numerical}, the model is numerically extended to the contact between self-affine rough surfaces, resulting in the description of three well-identified regimes of wear. A salient feature of the model is that the produced wear maps are function of well-defined physics-based model parameters, including material properties, surface roughness, and load.

\section{Analytical model for elastic interactions between micro-contacts}\label{sec:theory}

\begin{figure}
	\centering
	\includegraphics{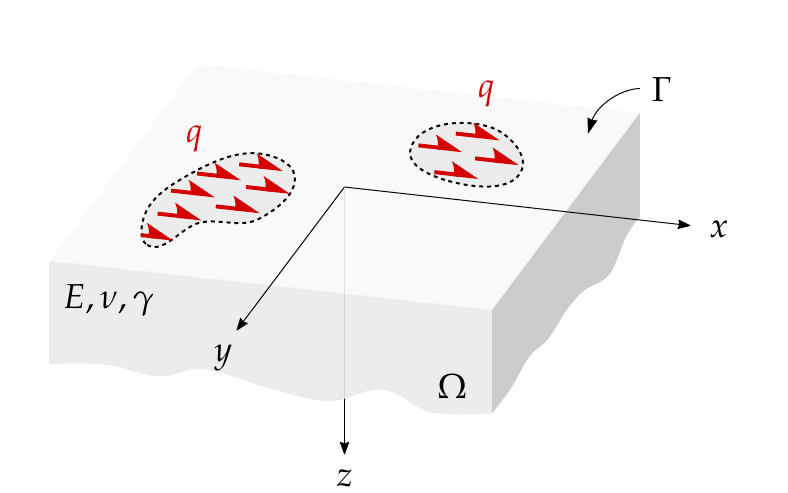}
	\caption{Micro-contacts under uniform tangential load on a semi-infinite solid}
	\label{fig:setup}
\end{figure}

Before diving into the contact between two rough surfaces, we take a look at a simplified contact between two flat surfaces, joined at a small number of cold-welded perfect junctions that we call micro-contacts. To study the elastic interaction between multiple micro-contacts during adhesive wear, we model them with uniform loads of magnitude $q$ acting along the $x$ direction at the surface $\Gamma$ of a semi-infinite solid $\Omega$ (see Figure~\ref{fig:setup}). Out of the two solids in contact, only the bottom one is considered because of symmetry.

In this section, adhesive wear is incorporated in this model by introducing two wear criteria, which are then used to find the definition of a critical size of micro-contact $d^*$ for the geometric configuration of Figure~\ref{fig:setup}. We then study analytically the interaction between two circular micro-contacts and produce a \emph{wear mechanism map}.

\subsection{Energy balance criterion}

\paragraph{Elastic energy}

When loaded, the solid $\Omega$ of Figure~\ref{fig:setup} accumulates elastic energy of deformation. If it is made of a linear elastic material, no energy is dissipated in the loading process, which implies that the elastic energy is equal to the work of the load on the surface:
\begin{equation}\label{eq:Eel3D}
	E_\text{el} = \frac{1}{2} \int_\Gamma \bm{u} \cdot \bm{p} \, d\Gamma \,,
\end{equation}
where $\bm{u}$ and $\bm{p}$ are respectively the displacement and traction fields on the surface $\Gamma$. In our case, the imposed tractions are only in the $x$ direction, so \eqref{eq:Eel3D} reduces to
\begin{equation}\label{eq:Eel}
	E_\text{el} = \frac{1}{2} \int_\Gamma u_x p_x \, d\Gamma \,.
\end{equation}
Here, the field $p_x$ describes the distribution of the tangential tractions in the $x$ direction on $\Gamma$, which are equal to $q$ wherever there is a micro-contact, and 0 otherwise (see Figure~\ref{fig:setup}). The surface displacements in the $x$ direction $u_x$ are determined from the fundamental solution established by Cerruti\cite{johnsonContactMechanics1985a}:
\begin{equation}\label{eq:ux_ker}
	u_{x\rightarrow x}^\text{ker} = \frac{1}{4\pi G} \left[ 2(1 - \nu)\frac{1}{r} + 2\nu\frac{x^2}{r^3} \right] \,,
\end{equation}
which is the displacement field in the $x$ direction caused by a unit point load at the origin of $\Omega$, also in the $x$ direction. $G$ is the shear modulus of the material, $\nu$ the Poisson's ratio, and $r$ is the distance from the origin: $r^2 = x^2 + y^2 + z^2$. The full displacement field $u_x$ is obtained by linear superposition of the contributions of all tractions:
\begin{align}
	u_x(x, y) &= \iint u_{x\rightarrow x}^\text{ker}(x - \xi, y - \eta) p_x(\xi, \eta) \, d\xi \, d\eta \\
	&= [u_{x\rightarrow x}^\text{ker} * p_x](x, y) \,, \label{eq:ux}
\end{align}
which is a convolution (denoted by the $*$ symbol). The tractions $p_x$ in the $x$ direction also induce displacements in the $y$ and $z$ direction, but they do not intervene in \eqref{eq:Eel}. We can rewrite \eqref{eq:Eel} as
\begin{equation}\label{eq:Eel_conv}
	E_\textnormal{el} = \frac{1}{2} \int_\Gamma [u_{x \rightarrow x}^\textnormal{ker} * p_x] \, p_x \, d\Gamma \,.
\end{equation}
Note that in order to be calculable analytically for simple cases, the convolution can be turned into a cross-correlation, as shown in Appendix~\ref{apx:conv_corr}.

\paragraph{Adhesive energy}

To detach a wear particle from $\Omega$, new surfaces have to be created, requiring adhesive energy (or fracture energy). We assume a simple spherical geometry. Therefore, the detachment of a single particle of diameter $d$ requires the creation of two hemispherical surfaces, needing an adhesive energy of
\begin{equation}
	E_\text{ad,1} = \pi\gamma d^2 \,. \label{eq:Ead}
\end{equation}

\paragraph{Criterion}

When a wear particle is detached from the bulk, it can no longer carry a tangential load applied at the surface. If a particle is detached where micro-contacts were present, those micro-contacts get unloaded and no longer contribute to the traction field $p_x$, resulting in a decrease $\Delta E_\text{el}$ of the elastic energy. This energy does not disappear, and in fact contributes to the formation of the cracks resulting in the particle detachment. Therefore, to determine if the particle can be fully detached, we consider the energy ratio
\begin{equation}
	\mathcal{R} = \frac{\Delta E_\text{el}}{E_\text{ad}} \,, \label{eq:R}
\end{equation}
and the particle can be detached if
\begin{equation}\label{eq:E_criterion}
	\mathcal{R} \geqslant 1 \,,
\end{equation}
which is the energy balance criterion.

\paragraph{Effect of normal load}

The creation of micro-contacts between two surfaces often results from the application of a normal load, which means that all the terms $p_x$, $p_z$, $u_x$ and $u_z$ contribute to the elastic energy \eqref{eq:Eel3D}. Nevertheless, the change $\Delta E_\text{el}$ in elastic energy due to unloading can still be solely attributed to the change of $p_x$ and $u_x$, neglecting the effect of the constant $p_z$. A proof is given in Appendix~\ref{apx:normal}.

\subsection{Crack initiation criterion}

Since the formation of wear particles results from the formation of cracks, their creation must start with the nucleation of such cracks, which can only be initiated at a point on a surface if
\begin{equation}
	\sigma_\text{I} \geqslant \sigma_\text{m} \,,
\end{equation}
where $\sigma_\text{I}$ is the first principal stress at this point, which is the maximum tensile stress if it is positive, and $\sigma_\text{m}$ is the tensile strength of the material.

\begin{figure}
	\centering
	\includegraphics{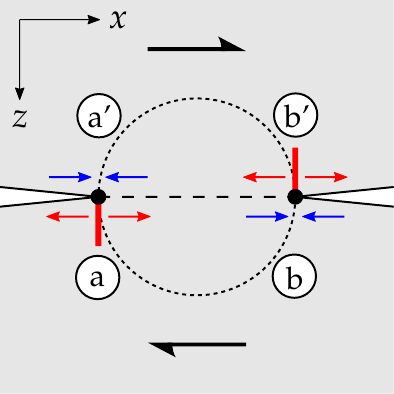}
	\caption[Cross section view of the required crack nucleation sites for the formation of a spherical wear particle at a micro-contact]{Cross section view of the required crack nucleation sites for the formation of a spherical wear particle at a micro-contact. Points (a) and (b) are slightly below the junction, and points (a') and (b') slightly above. The thick red lines show the cracks which must be nucleated in order to detach the particle. Tensile stresses (red arrows) must overcome the tensile strength of the material. Equal and opposite compressive stresses (blue arrows) appear by symmetry.}
	\label{fig:crack_ini_crit}
\end{figure}

Until now, we only considered the bottom solid to simplify the problem. Let us also reconsider the top solid for a moment. In order to detach a spherical wear particle, two diametrically opposed cracks have to be initiated, as shown in Figure~\ref{fig:crack_ini_crit} as thick red lines at locations (a) in the bottom solid and (b') in the top one. The cracks can be nucleated if the tensile stress at those points, shown by red arrows, is sufficiently large. Thanks to the symmetry of the loading, we can state that the maximum tensile stress at the point (b') in the top solid is equal in magnitude to the maximum compressive stress at the point (b) in the bottom solid, so that the conditions for crack opening can be defined by looking only at the bottom solid. In summary, when considering only the bottom solid, a particle can be detached if it has a sufficiently large tensile stress at its trailing edge (a) and a sufficiently large compressive stress at its leading edge (b). The crack initiation criterion can therefore be written as:
\begin{subequations}\label{eq:crack_ini}
\begin{align}
	\sigma_\text{I} \geqslant \sigma_\text{m} \quad \text{at point (a)} \,, \\
	\sigma_\text{III} \leqslant -\sigma_\text{m} \quad \text{at point (b)} \,.
\end{align}
\end{subequations}

In the way our micro-contacts are defined (Figure~\ref{fig:setup}), the tangential traction field at the interface oriented in the $x$ direction has discontinuities from 0 to $q$ at the borders of the micro-contacts, which leads to stress singularities (regions of infinite stresses) in the other directions inside the solid around those places\cite{johnsonContactMechanics1985a}, also leading to infinite principal stresses. Therefore, the crack initiation criterion \eqref{eq:crack_ini} can always be satisfied somewhere on the borders of the micro-contacts.

\subsection{Single sheared micro-contact and critical size}

In order to define a critical junction size as in Aghababaei \emph{et al.}\cite{aghababaeiCriticalLengthScale2016}, we  consider a single circular micro-contact of diameter $d$. The stored elastic energy stored by tangentially loading the micro-contact can be calculated analytically from \eqref{eq:Eel_conv} (see Appendix~\ref{apx:Eel1}), leading to
\begin{equation}\label{eq:Eel1}
	E_\text{el,1} = \frac{(2 - \nu) d^3 q^2}{12G} \,.
\end{equation}
The corresponding adhesive energy required to detach a hemispherical wear particle under this circular micro-contact is given in \eqref{eq:Ead}. From these two expressions, we obtain the energy ratio from \eqref{eq:R}:
\begin{equation}
	\mathcal{R} = \frac{(2 - \nu) d q^2}{12\pi \gamma G} \,.
\end{equation}

The maximum tangential load $q$ which can be applied on the micro-contact is equal to the shear strength $\sigma_\text{j}$ of the junction between the two surfaces in contact. After setting $q = \sigma_\text{j}$ in the expression of $\mathcal{R}$, we look for the value of $d$ which makes $\mathcal{R}$ satisfy the energy balance criterion \eqref{eq:E_criterion}. We find a critical diameter:
\begin{equation}\label{eq:d_star}
	d^* = \frac{12\pi\gamma G}{(2 - \nu)\sigma_\text{j}^2}
\end{equation}
which only depends on material parameters. The energy balance criterion is satisfied whenever $d \geqslant d^*$\snspace. Note that the expression of $d^*$ found in this geometrical configuration is in accordance with the expression found by Aghababaei \emph{et al}. \eqref{eq:d_star_ramin} with a newly defined geometrical factor.

A single sheared micro-contact of diameter $d$ can result in the detachment of a wear particle if $d \geqslant d^*$ and otherwise flows plastically.

\subsection{Interaction of two micro-contacts}

\begin{figure}
	\centering
	\includegraphics{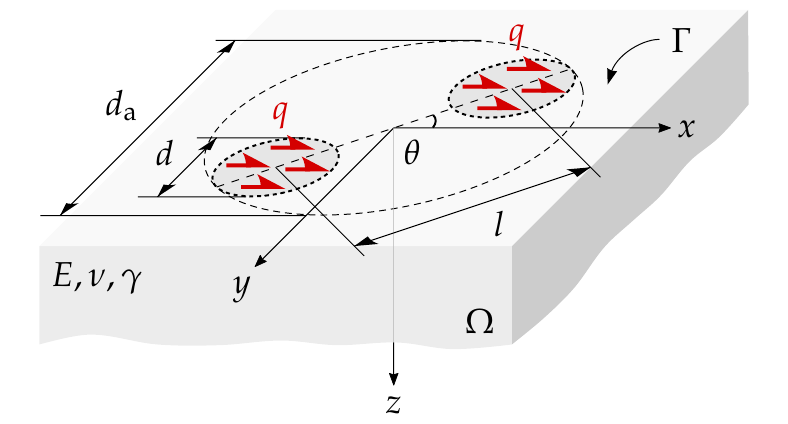}
	\caption[Two circular contacts under uniform tangential load]{Two circular contacts under uniform tangential load. $d_\text{a}$ is the `apparent' diameter of a hemispherical particle that would encompass both micro-contacts, and $\theta$ is the angle between the direction of the load and the line going through the centers of the micro-contacts.}
	\label{fig:2_circular_contacts}
\end{figure}

To study the elastic interactions between multiple micro-contacts in the same manner as the two-dimensional model of Pham-Ba \emph{et al.}\cite{pham-baAdhesiveWearInteraction2020}, we start by considering two tangentially loaded circular micro-contacts of diameters $d$ and having a distance $l$ between their centers, as shown in Figure~\ref{fig:2_circular_contacts}. They are tangentially loaded in the $x$ direction with a pressure of magnitude $\sigma_\text{j}$ and the line going through both of their centers makes an angle $\theta$ with the $x$ axis.

\paragraph{Elastic energy}

\begin{figure} % From 9.5 notebook
	\centering
	\includegraphics{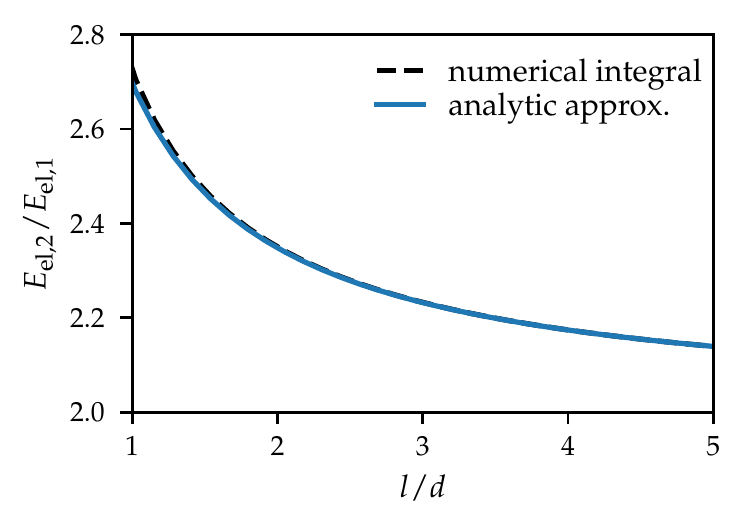}
	\caption[Comparison of the numerical integration and the analytical approximation of the elastic energy stored under two tangentially loaded circular contacts]{Comparison of the numerical integration and the analytical approximation of the elastic energy stored under two tangentially loaded circular contacts. Here, $\nu = 0.3$ and $\theta = 0$. The analytical approximation is more accurate when $l$ is large.}
	\label{fig:Eel2}
\end{figure}

The calculation of the stored elastic energy in this case requires rewriting \eqref{eq:Eel_conv} to be calculable by hand. A good approximation of the elastic energy is derived in Appendix~\ref{apx:Eel2}:
\begin{equation}\label{eq:Eel2}
	E_\text{el,2} \approx \frac{(2 - \nu) d^3 \sigma_\text{j}^2}{6G} + \frac{\pi d^4\sigma_\text{j}^2}{32G} \frac{1 - \nu \sin^2\theta}{l} \,,
\end{equation}
which is exact in the limit when $l \gg d$ but is still very accurate when $l$ reaches $l = d$, which is when the two micro-contacts are adjacent. The accuracy is verified by numerically integrating \eqref{eq:Eel_conv} for fixed values of $d$ and $l$ and comparing the results with \eqref{eq:Eel2}, as shown in Figure~\ref{fig:Eel2}.

The expression \eqref{eq:Eel2} consists in the sum of two terms. The left term is equal to $2 E_\text{el,1}$ \eqref{eq:Eel1} and represents the energetic contributions of each circular micro-contact. The right terms is proportional to $1/l$ and represents the effect of the elastic interactions between the two micro-contacts. When $l$ is large compared to $d$, the right term vanishes and the two micro-contacts do not interact: the total energy is just the sum of their energies taken separately. When $l$ decreases, the two micro-contacts get closer and the interaction term increases (see Figure~\ref{fig:Eel2}).

Note that if the two micro-contacts were to be superimposed into a single circular micro-contact loaded at $2q$, the resulting elastic energy would reach $4E_\text{el,1}$, as \eqref{eq:Eel1} has a quadratic dependence on $q$. This would not happen in practice when $q = \sigma_\text{j}$ since the tangential load is limited to $\sigma_\text{j}$, but it explains why the $E_\text{el,2}$ is necessarily bounded between $2E_\text{el,1}$ and $4E_\text{el,1}$.

\paragraph{Adhesive energy}

\begin{figure}
	\centering
	\vspace{-0.5cm}
	\subfloat[One separated]{
		\includegraphics{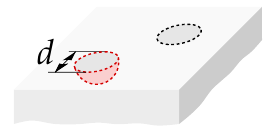}
	}
	\subfloat[Two separated]{
		\includegraphics{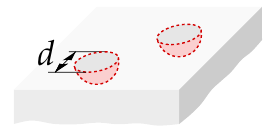}
	}
	\subfloat[Combined]{
		\includegraphics{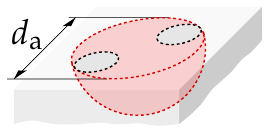}
	}
	\caption[Different cases of wear particle formation with two micro-contacts]{Different cases of wear particle formation with two micro-contacts. The red surfaces show the surfaces created when the particle is detached. The case where no wear particle is detached is not shown.}
	\label{fig:2_mc_particles}
\end{figure}

With two loaded micro-contacts, several cases of wear particle formation may arise, as shown in Figure~\ref{fig:2_mc_particles}. There can be either zero, one or two separated wear particles at each micro-contact, or a single combined wear particle encompassing both micro-contacts. We call $d_\text{a}$ (for \emph{apparent} diameter) the diameter of the potentially formed combined wear particle. We have $d_\text{a} = d + l$.

In the case of the formation of a single separated particle,\footnote{When there is enough energy to form only a single particle, a slight asymmetry in the system would select one of the two contact junctions and the other junction would deform plastically.} the required adhesive energy to detach the particle is the same as \eqref{eq:Ead}:
\begin{equation}
	E_\text{ad,1sep} = \pi\gamma d^2 \,. \label{eq:Ead1sep}
\end{equation}
In the case of the formation of two wear particles, the required adhesive energy is twice as big:
\begin{equation}
	E_\text{ad,2sep} = 2\pi\gamma d^2 \,, \label{eq:Ead2sep}
\end{equation}
and in the case of the formation of a combined wear particle, it is
\begin{equation}
	E_\text{ad,comb} = \pi\gamma d_\text{a}^2 \,. \label{eq:Eadcomb}
\end{equation}

\paragraph{Energy balance criterion}

Assuming that both micro-contacts get unloaded when the two separated wear particles or a single combined one are formed, the decrease of elastic energy $\Delta E_\text{el}$ in \eqref{eq:R} is equal to $E_\text{el,2}$ \eqref{eq:Eel2}. From the expression of $\Delta E_\text{el}$ and the different expressions of the adhesive energy for each case, we get the energy ratio
\begin{equation}\label{eq:R2}
	\mathcal{R} = \left( C_\text{n} + \frac{3\pi}{8} C_\text{inter} \frac{d}{l} \right ) \frac{d}{d^*} C_\text{case}
\end{equation}
where $C_\text{n} = 2$, and
\begin{equation}
	C_\text{inter} = \frac{1-\nu\sin^2\theta}{2-\nu}
\end{equation}
is a constant controlling the amount of interaction, which only depends on the Poisson's ratio and the angle $\theta$ and has always a value between $1/3$ and $2/3$. Choosing $\nu = 0.5$ and $\theta = 0$ (micro-contacts aligned with the direction of the load) provides the most interaction with $C_\text{inter} = 2/3$, whereas $\theta = \pi / 2$ (line of the micro-contacts perpendicular to the direction of the load) gives the least amount of interaction with $C_\text{inter} = 1/3$. We will discuss the implications of this in the next sub-section. The constant $C_\text{case}$ has a value which depends on the case of particle formation. From the different adhesive energies $E_\text{ad,2sep}$ \eqref{eq:Ead2sep} and $E_\text{ad,comb}$ \eqref{eq:Eadcomb}, we have $C_\text{case} = 1/2$ or $C_\text{case} = d^2/d_\text{a}^2$ respectively. Note that the material parameters $G$, $\sigma_\text{j}$ and $\gamma$ do not appear directly in \eqref{eq:R2} and were conveniently replaced by the critical diameter $d^*$ \eqref{eq:d_star} which contains all those missing terms.

When only one micro-contact out of the two forms a wear particle and gets unloaded, the elastic energy goes from $E_\text{el,2}$ \eqref{eq:Eel2} to $E_\text{el,1}$ \eqref{eq:Eel1} with $q = \sigma_\text{j}$, as the remaining micro-contact, flowing plastically, still carries a load of $\sigma_\text{j}$. Therefore, the decrease of elastic energy is $\Delta E_\text{el} = E_\text{el,2} - E_\text{el,1}$, which with the expression of the adhesive energy $E_\text{ad,1sep}$ \eqref{eq:Ead1sep} gives the energy ratio \eqref{eq:R2} with $C_\text{n} = 1$ and $C_\text{case} = 1$.

\subsection{Wear map}

When we derived the energy ratio for a single micro-contact and the formation of a single wear particle \eqref{eq:R}, we were able to find a critical diameter $d^*$ which easily defines which behaviour is expected (plastic flow or formation of a particle). The expressions of the energy ratio $\mathcal{R}$ for two micro-contacts \eqref{eq:R2} are more complicated, as they now depend on $d$, $l$, $d^*$ and $C_\text{inter}$.

\begin{figure} % From 10.3.2 notebook
	\centering
	\includegraphics{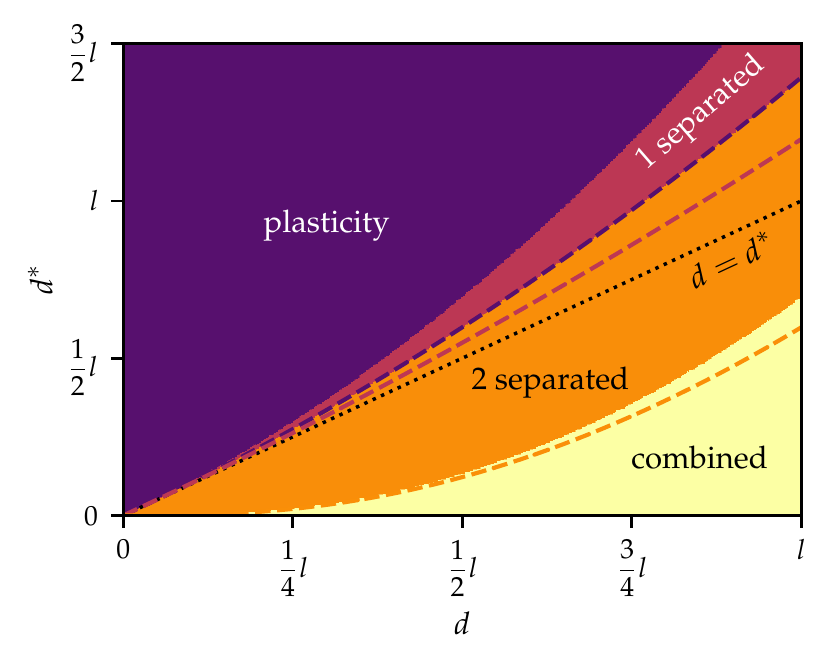}
	\caption[Wear map of the different cases of wear particle formation under tangentially loaded two micro-contacts]{Wear map of the different cases of wear particle formation under tangentially loaded two micro-contacts. The colored regions are computed at maximum possible elastic interaction ($\nu = 0.5$, $\theta = 0$). The dashed colored lines show the boundaries the regions would have at minimum interaction ($\nu = 0.5$, $\theta = \pi/2$).}
	\label{fig:wear_map_2mc}
\end{figure}

Figure~\ref{fig:wear_map_2mc} shows our best attempt to represent the different possible scenarii of wear particle formation as a function of model parameters. We refer to Figure~\ref{fig:wear_map_2mc} as a \emph{wear map}. Each colored region shows where $\mathcal{R} \geqslant 1$ for a selected value of $C_\text{case}$ and therefore tells that the indicated behaviour is energetically feasible, depending on the contact junction size $d$ and the critical junction size $d^*$\snspace, both expressed relative to $l$. The colored regions in Figure~\ref{fig:wear_map_2mc} are computed at maximum interaction with $C_\text{inter} = 2/3$. The dashed colored lines show the boundaries that those regions would have if $C_\text{inter} = 1/3$, that is at minimum interaction. Clearly, increasing elastic interactions favors wear particle formation.

The black dotted line $d = d^*$ in Figure~\ref{fig:wear_map_2mc} is given for comparison with the single micro-contact case. Indeed, such system would be in a plastic regime when $d < d^*$\snspace, which is the whole region above the dotted line, and it would allow the formation of a wear particle when $d \geqslant d^*$\snspace, which is the region below the line. The wear map predicts that with two interacting micro-contacts, it is possible to form one and even two wear particles, even if the size of each micro-contact is smaller than the minimum required $d^*$\snspace. Once more, we emphasize that this is because of elastic interactions, which make the available stored elastic energy larger that trivially expected, as explained by \eqref{eq:Eel2}.

The wear map hints toward the emergence of multiple wear regimes. The `plasticity' region corresponds to theoretically no wear volume, as the sliding surfaces only get plastically remodeled. This behavior can be linked to the experimental observations of low wear. In contrast, the jump in wear volume between the `2 separated' and the `combined' regions is significant, and can be related to the transition between a mild and a severe wear regime.

It is possible to derive wear maps for larger numbers of micro-contacts and other arrangements, following the same derivation for the analytical approximation of the elastic energy (Appendix~\ref{apx:Eel2}). However, we will show in the next section that more interesting results can be obtained using numerical simulations with generic rough surfaces while following the same energetic principles.

\section{Numerical model for random rough surfaces}\label{sec:numerical}

In the previous section, we considered systems with one or two circular micro-contacts of a given diameter and with a simple parameterized arrangement on the surface. In reality, the micro-contacts between two loaded nominally flat surfaces emerge from random surface roughness at lower scales, and therefore come in various shapes and sizes.  We resort to numerical simulations to describe them accurately.

\subsection{Description}

\paragraph{Distribution of micro-contacts}

\begin{figure}
	\centering
	\includegraphics{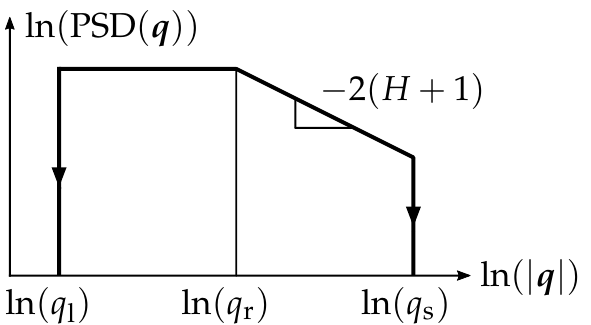}
	\caption[Target power spectral density of generated rough surfaces]{Target power spectral density of generated rough surfaces. Its value is zero at frequencies $q < q_\text{l}$ and $q > q_\text{s}$. $H$ is the Hurst exponent, $q_\text{l}$, $q_\text{r}$, and $q_\text{s}$ are respectively the frequencies corresponding to the largest, roll-off and smallest wavelengths.}
	\label{fig:psd}
\end{figure}

We use the open source software \emph{Tamaas}\cite{frerotTamaasLibraryElasticplastic2020} to generate discretized self-affine\cite{perssonContactMechanicsRandomly2006} rough surfaces $h(x, y)$ with the power spectral density\footnote{It is defined as the squared magnitude of the Fourier transform of $h$.} shown in Figure~\ref{fig:psd}. The rough surfaces are discretized on a grid of $n \times n$ points, with $n = 512$, and of side-length $L$. Their parameters are the root mean square (RMS) of slopes $\sqrt{\langle|\nabla h|^2\rangle}$, the Hurst exponent $H$ and the frequencies $q_\text{l}$ and $q_\text{s}$ corresponding respectively to the largest and smallest wavelengths contained in the spectrum, where a frequency of the surface is given by $q = k\frac{2\pi}{L}$, also called wavenumber, and $k$ is a number ranging from $0$ to $\lceil\frac{n}{2}\rceil$. We set the roll-off frequency $q_\text{r}$ at $q_\text{r} = q_\text{l}$.

\begin{figure*}[p]
	\centering
	\subfloat[$\tilde{p}_\text{N} = 0.02$]{
		\includegraphics[width=3.9cm]{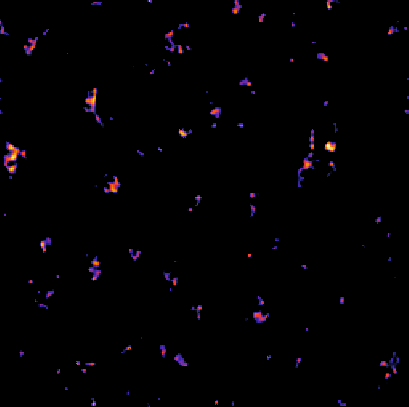}
	}
	\subfloat[$\tilde{p}_\text{N} = 0.05$]{
		\includegraphics[width=3.9cm]{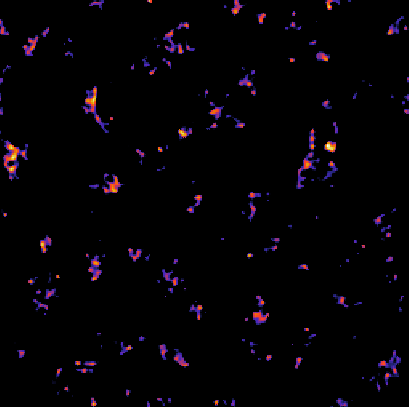}
	}
	\subfloat[$\tilde{p}_\text{N} = 0.1$]{
		\includegraphics[width=3.9cm]{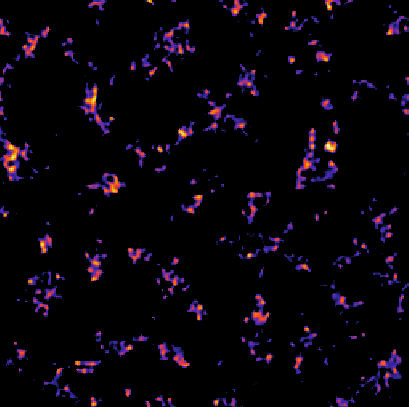}
	}
	\subfloat[$\tilde{p}_\text{N} = 0.2$]{
		\includegraphics[width=3.9cm]{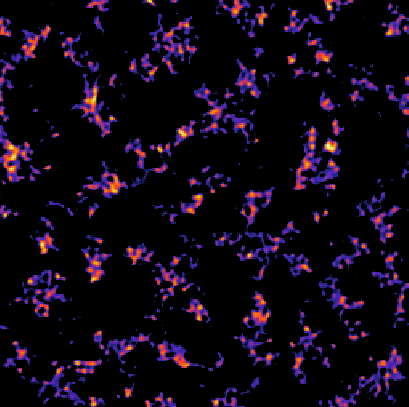}
	}
	\caption[Boundary element simulations of the micro-contacts and local contact pressures for increasing normal load]{Boundary element  simulations of the micro-contacts and local contact pressures for increasing normal load. The normalized normal load is indicated, which also corresponds to the ratio of real contact area to apparent contact area. Brighter color corresponds to a higher local pressure. For increasing normal load, the number and size of the micro-contacts increase, until micro-contacts become large enough to merge together, resulting in a drop in the number of contacts but a sharp rise of their average size. The surface roughness parameters are $n = 512$, $H = 0.8$, $q_\text{l} = 8$ and $q_\text{s} = n/8$.}
	\label{fig:rough_contact}
\end{figure*}

\emph{Tamaas}, primarily a boundary element method software, is used to efficiently solve the elastic contact between two rough surfaces, equivalently considered as the contact between a rigid rough surface (with equivalent roughness) and a flat deformable elastic solid. Figure~\ref{fig:rough_contact} shows distributions of the contact pressure on a rough surface with $H = 0.8$, $q_\text{l} = 8$ and $q_\text{s} = n/8$ for different normal loads. We normalize the normal load\cite{hyunFiniteelementAnalysisContact2004}:
\begin{equation}
	\tilde{p}_\text{N} = \frac{\sqrt{2\pi}}{\sqrt{\langle|\nabla h|^2\rangle}}\frac{p_\text{N}}{E^*} \,,
\end{equation}
where $p_\text{N}$ is the normal load and
\begin{equation}
	E^* = \frac{E}{1 - \nu^2}
\end{equation}
is the effective Young's modulus. At low normal loads, the normalized normal load $\tilde{p}_\text{N}$ is a good approximation of the ratio between the real and the apparent contact area.

For given rough surface parameters $q_\text{l}$, $q_\text{s}$ and $H$, the normalized load $\tilde{p}_\text{N}$ is the only free parameter for the description of the rough contact, and it combines the effect of the normal load and the RMS of heights. In the contact simulations, all the grid points where the local normal pressure is non-zero are in contact. They give the needed locations of micro-contacts.

\paragraph{Detachment of wear particles}

We use the energy balance and the crack initiation criteria on the distribution of micro-contacts to determine the potential wear particle formation sites. Assuming a constant tangential load $\sigma_\text{j}$ in the contact areas, the elastic energy can be numerically computed with \eqref{eq:Eel_conv}. The procedure to find the maximum wear volume is the following:
\begin{enumerate}
	\item Consider largest particle fulfilling crack initiation criterion;
	\item Unload corresponding region $\mathcal{C}$ (remove tangential loads) and compute the drop in elastic energy $\Delta E_\text{el}$;
	\item If $\Delta E_\text{el}$ greater than needed $E_\text{ad}$ (for a given $d^*$), save this particle removal; Else, try next largest particle;
	\item Repeat until no more particles can be added.
\end{enumerate}
After those steps, the remaining elastic energy should be small and insufficient to allow the creation of further particles. The position and size of each created particle is recorded for analysis.

As a result of this procedure, a list of wear particles (position and size) is obtained for any given value of $\tilde{p}_\text{N}$ (controlling the micro-contacts) and $d^*$ (controlling the ductile to brittle transition).

Note that this algorithm requires many consecutive explicit calculations of $\Delta E_\text{el}$ for testing the unloading of each possible particle (according to the crack initiation criterion) and is computationally expensive. Indeed, the creation of a particle unloads a portion $\mathcal{C}$ of the domain $\Omega$, affecting the displacements, going from $u_x$ to $u_x'$ (computed with \eqref{eq:ux} for each possible $\mathcal{C}$), with $u_x' < u_x$ at every point of $\Omega$ (by the principle of superposition, less tractions are applied after unloading). The resulting expression of elastic energy release is therefore:
\begin{align}
	&\Delta E_\text{el} = \frac{1}{2}\int_{\Omega} p_x\, u_x\, d\Omega - \frac{1}{2}\int_{\Omega\setminus\mathcal{C}} p_x\, u_x'\, d\Omega \nonumber \\
	&= \frac{1}{2}\int_{\mathcal{C}} p_x\, u_x\, d\Omega + \frac{1}{2} \left( \int_{\Omega\setminus\mathcal{C}} p_x\, u_x\, d\Omega - \int_{\Omega\setminus\mathcal{C}} p_x\, u_x'\, d\Omega \right) \nonumber \\
	&= \frac{1}{2}\int_{\mathcal{C}} p_x\, u_x\, d\Omega + \frac{1}{2} \int_{\Omega\setminus\mathcal{C}} p_x\, (u_x - u_x') \, d\Omega \,. \label{eq:DEel}
\end{align}
In comparison, to find at each iteration a detached particle, the approach of Popov and Pohrt\cite{popovAdhesiveWearParticle2018} estimates the energy release by integrating the local elastic energy density, without computing $u_x'$. In this case, the drop in elastic energy takes the form:
\begin{equation*}
	\Delta E_\text{el} = \frac{1}{2}\int_{\mathcal{C}} p_x\, u_x\, d\Omega \,,
\end{equation*}
which misses the additionnal positive term present in \eqref{eq:DEel}. While less computationnaly intensive, the procedure adopted by Popov and Pohrt leads to an underestimation of elastic interations. As a consequence, a single power law between wear volume and applied pressure is found. There is no visible transition to severe wear, which only arises with proper accounting of elastic interactions.

\subsection{Validation}

\begin{figure*} % From 11.2.2 notebook
	\centering
	\subfloat[Plasticity]{
		\includegraphics[width=39mm, trim=30 30 70 30, clip]{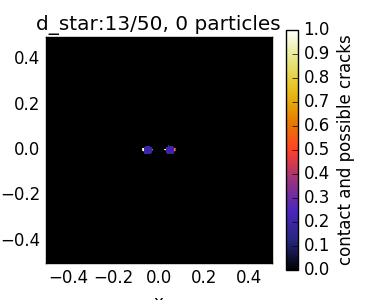}
	}
	\subfloat[1 separated]{
		\includegraphics[width=39mm, trim=30 30 70 30, clip]{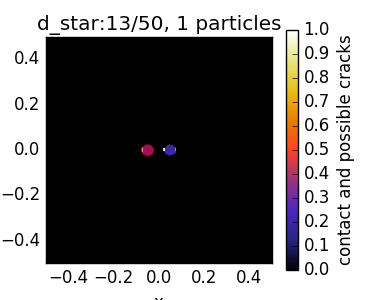}
	}
	\subfloat[2 separated]{
		\includegraphics[width=39mm, trim=30 30 70 30, clip]{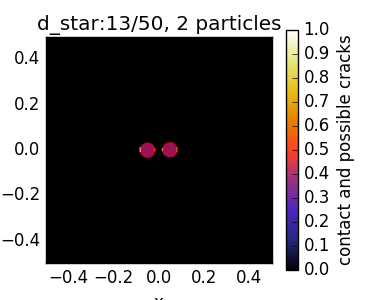}
	}
	\subfloat[Combined]{
		\includegraphics[width=39mm, trim=30 30 70 30, clip]{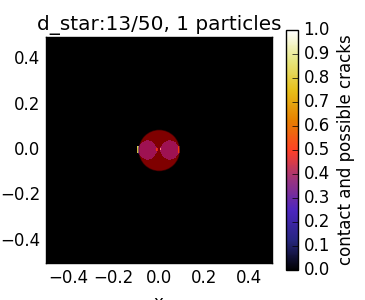}
	}
	\caption[Different cases of wear particle formation with two micro-contacts in the numerical model]{Different cases of wear particle formation with two micro-contacts in the numerical model. The blue regions are the micro-contacts. The red circles are the wear particles which can be detached. The small brighter spots are the regions of tensile and compressive stresses where the crack initiation criterion can be fulfilled.}
	\label{fig:val_particles}
\end{figure*}

The computation of the detachment of wear particles in the numerical model was tested with a setup consisting of two micro-contacts, for which we previously derived an analytical theory and a constructed a wear map. We are choosing the micro-contacts to be aligned with the load ($\theta=0$). Figure~\ref{fig:val_particles} shows the effect of increasing $d$ for constant values of $l$ and $d^*$\snspace. Looking at the wear map (Figure~\ref{fig:wear_map_2mc}), this means moving on an horizontal line from left to right. As predicted by the wear map, there is a transition between all the possible behaviors of wear particle formation.

\begin{figure*}
	\centering
	\subfloat[Wear map]{
		\includegraphics{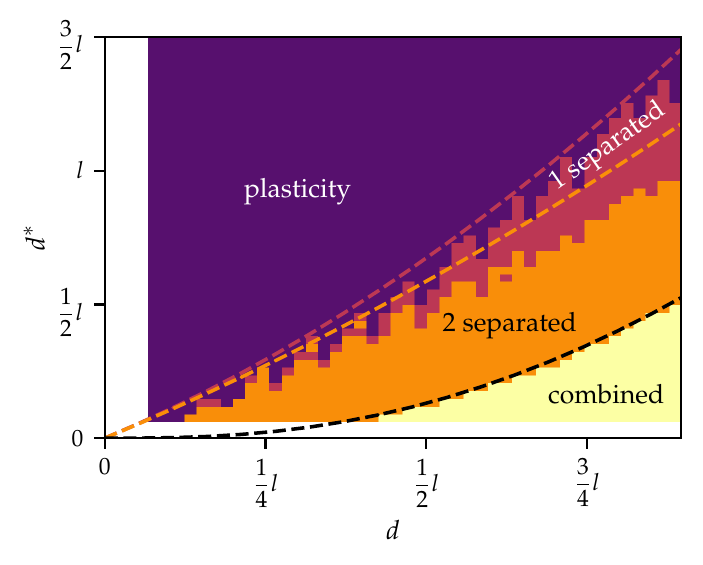}
	}
	\subfloat[Wear volume]{
		\includegraphics{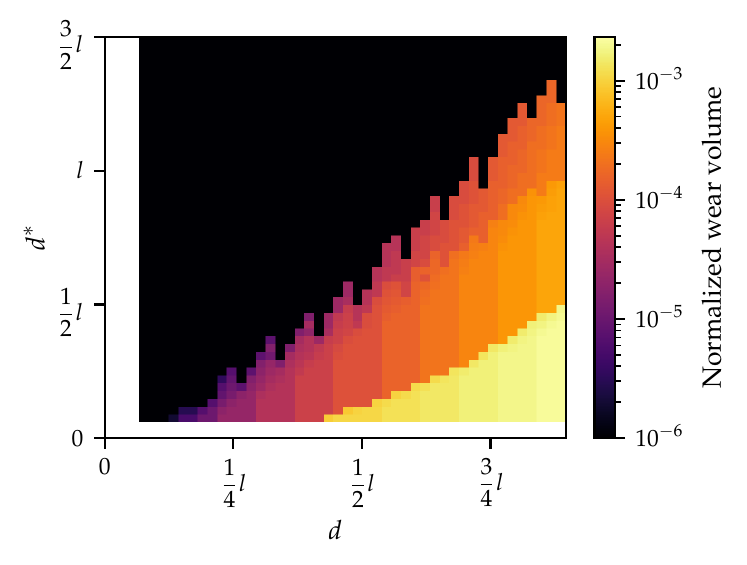}
	}
	\caption[Numerical wear maps for two circular micro-contacts]{Numerical wear maps for two circular micro-contacts. Here, $\theta = 0$ and $\nu = 0.3$. The jagged edges of the regions are due to the coarse discretization of the simulated surface. \textbf{(a)}~Wear map deduced from the number of formed particles (0, 1 or 2) and their volume. The four regions of the analytical wear map (Figure~\ref{fig:wear_map_2mc}) are recovered, and the analytical boundaries are shown with dashed lines. \textbf{(b)}~The wear volume is normalized by $L^3$, where $L$ is the side-length of the discretization surface.}
	\label{fig:val_wear_maps}
\end{figure*}

The numerically generated wear maps are shown in Figure~\ref{fig:val_wear_maps}. They are generated by varying the values of $d$ and $d^*$ and by computing the possible detachment of wear particles. Figure~\ref{fig:val_wear_maps}(a) shows the different behaviours deduced by the number of formed particles and their volume. It agrees with the analytical wear map (Figure~\ref{fig:wear_map_2mc}) superimposed by dashed lines.  Figure~\ref{fig:val_wear_maps}(b) indicates the corresponding wear volumes. It shows that in the lower right region of the wear map, a much higher wear volume is created, which corresponds to the `combined' behavior of particle formation. Note that the transition to this behavior is quite sharp.

\subsection{Wear maps and wear regimes}

\begin{figure*}
	\centering
	\subfloat[$\frac{d^*}{L} = 0.13$, $\tilde{p}_\text{N} = 0.02$]{
		\includegraphics[width=39mm, trim=30 30 70 30, clip]{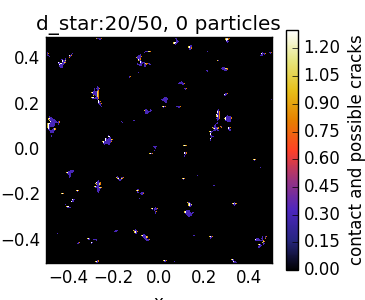}
	}
	\subfloat[$\frac{d^*}{L} = 0.13$, $\tilde{p}_\text{N} = 0.05$]{
		\includegraphics[width=39mm, trim=30 30 70 30, clip]{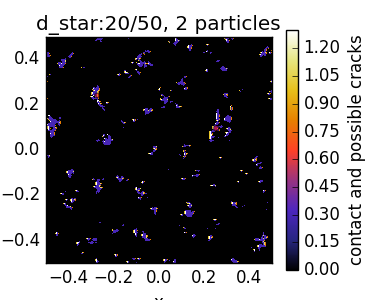}
	}
	\subfloat[$\frac{d^*}{L} = 0.13$, $\tilde{p}_\text{N} = 0.1$]{
		\includegraphics[width=39mm, trim=30 30 70 30, clip]{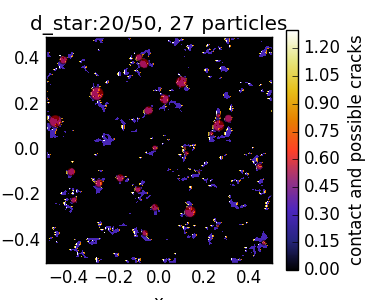}
	}
	\subfloat[$\frac{d^*}{L} = 0.13$, $\tilde{p}_\text{N} = 0.2$]{
		\includegraphics[width=39mm, trim=30 30 70 30, clip]{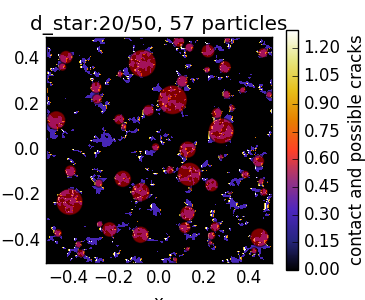}
	}
	\\
	\subfloat[$\frac{d^*}{L} = 0.04$, $\tilde{p}_\text{N} = 0.02$]{
		\includegraphics[width=39mm, trim=30 30 70 30, clip]{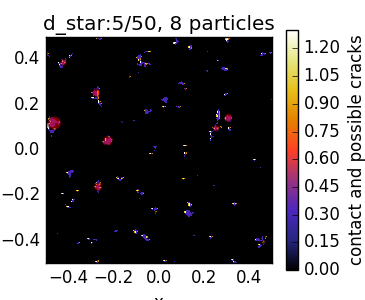}
	}
	\subfloat[$\frac{d^*}{L} = 0.04$, $\tilde{p}_\text{N} = 0.05$]{
		\includegraphics[width=39mm, trim=30 30 70 30, clip]{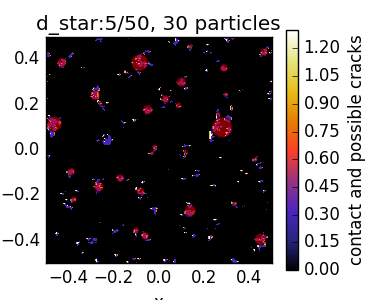}
	}
	\subfloat[$\frac{d^*}{L} = 0.04$, $\tilde{p}_\text{N} = 0.1$]{
		\includegraphics[width=39mm, trim=30 30 70 30, clip]{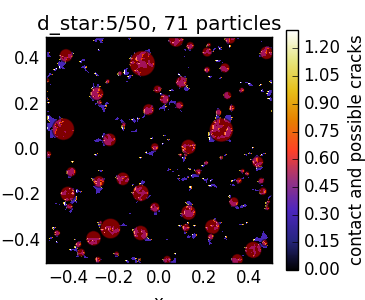}
	}
	\subfloat[$\frac{d^*}{L} = 0.04$, $\tilde{p}_\text{N} = 0.2$]{
		\includegraphics[width=39mm, trim=30 30 70 30, clip]{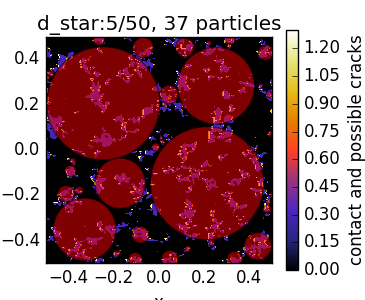}
	}
	\caption[Different cases of wear particle formation in a rough contact]{Different cases of wear particle formation in a rough contact. \textbf{(a)}-\textbf{(d)}~When the normal load increases, the number and size of wear particles increase, following the trend of the micro-contacts. \textbf{(e)}-\textbf{(h)}~With a lower $d^*$\snspace, particles are generated at lower loads. Also, at high loads, elastic interactions promote the formation of less numerous and larger particles encompassing multiple micro-contacts, even if the distributions of micro-contacts are the same as above.}
	\label{fig:rough_particles}
\end{figure*}

\begin{figure*}
	\centering
	\subfloat[Number of particles]{
		\includegraphics{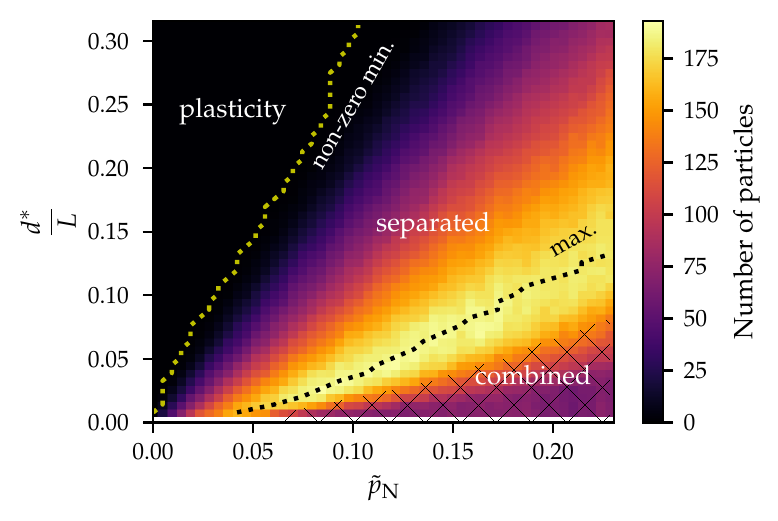}
	}
	\subfloat[Wear volume]{
		\includegraphics{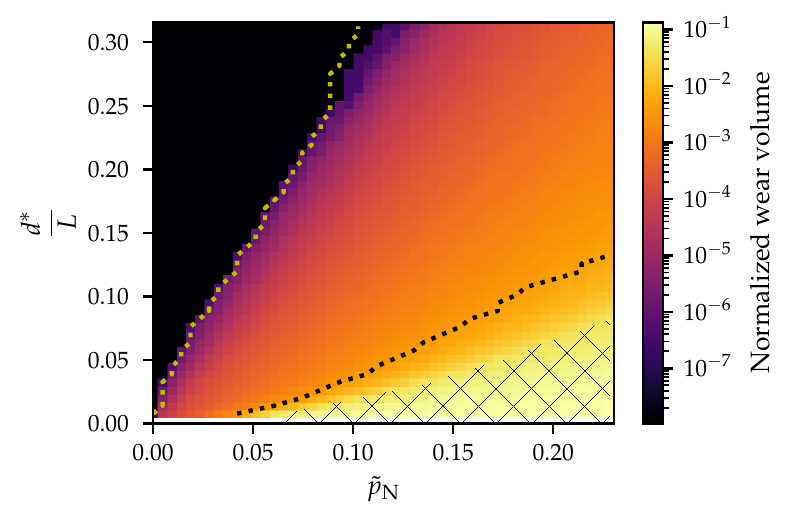}
	}
	\caption[Wear maps of the contact between rough surfaces]{Wear maps of the contact between rough surfaces. The map of the number of particles shows clearly distinct regions. Between the `separated' and `combined' regions, the number of particles decreases but the wear volume increases. The wear volume spans multiple orders of magnitude. The crossed area is the region where the numerical simulation validity is not guaranteed, because the size of the wear particles becomes comparable to the size of the simulated system.}
	\label{fig:rough_wear_maps}
\end{figure*}

The numerical model was run on five randomized rough surfaces with the roughness parameters $H = 0.8$, $q_\text{l} = 8$ and $q_\text{s} = n/8$ (with $n = 512$). Examples of wear particle distributions are shown in Figure~\ref{fig:rough_particles} and the averaged computed wear maps are given in Figure~\ref{fig:rough_wear_maps}. Regions similar to the ones of the wear map for two micro-contacts can be found. The `plasticity' region is where no wear particles are detached. Then, for a constant $d^*$ (\emph{i.e.} for a given material), the number of particles increases with the load until reaching a maximum value, defining a region `separated' where separated particles can be formed, as shown in Figure~\ref{fig:rough_particles}(a) to (g). Then, the number of particles decreases with the higher loads, entering the `combined' region, where large wear particles can encompass multiple micro-contacts. The wear volume increases monotonically with the load and reaches a plateau (the crossed areas in the wear maps), which is a non-physical numerical artifact caused by the fact that wear particles reach the size of the discretized system.

The effects of the material parameters can also be read on the wear maps, since they are contained into $d^*$\snspace. According to our model, a material with a lower $d^*$\snspace, that would be harder or more fragile, should form smaller particles. The model also predicts that harder materials are more prone to generate combined particles from neighboring contact junctions.

However, the full volume of debris production is higher for a harder material, which seems in opposition to Archard's wear law\cite{archardContactRubbingFlat1953}. This is a limitation of not accounting for the sliding history, as we discuss further below. It is also a consequence of assuming that our junctions all carry the material specific shear strength $\sigma_\text{j}$, implying that harder materials are loaded tangentially with a larger force. As more mechanical work is imparted to the interface for hard materials, this results in larger wear volume production. The exact distributions of shear forces at micro-contacts should be examined in future work.

\begin{figure*}
	\centering
	\subfloat[Number of particles]{
		\includegraphics{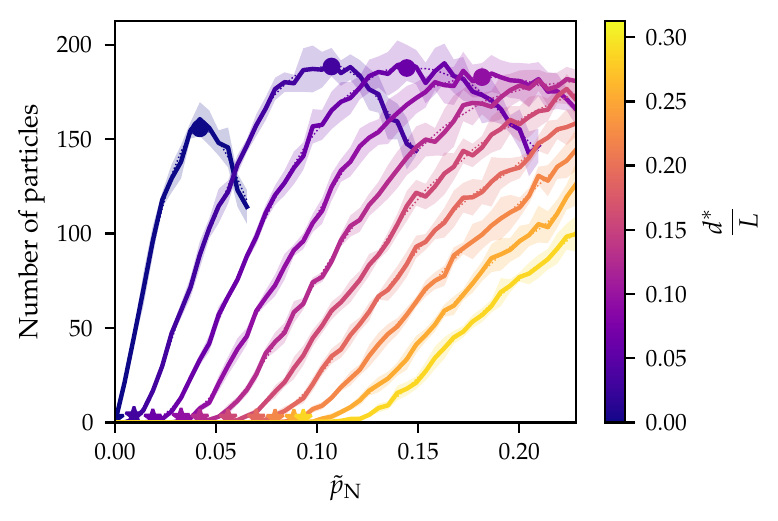}
	}
	\subfloat[Wear volume]{
		\includegraphics{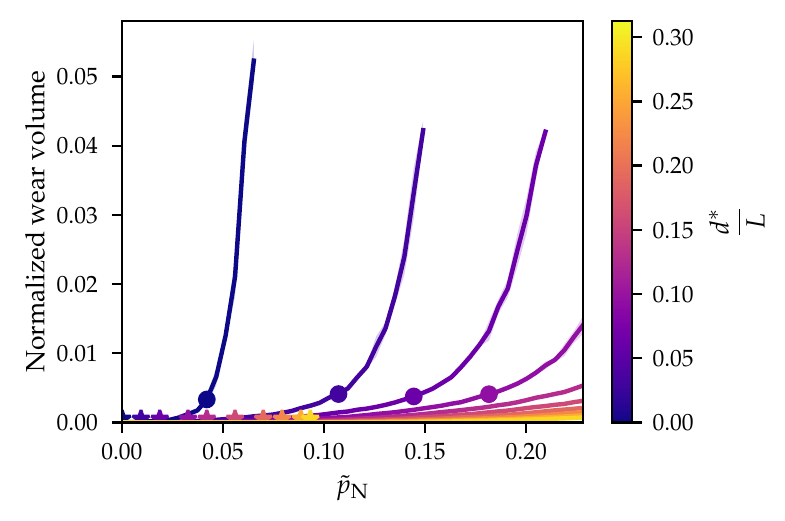}
	}
	\caption[Wear curves of the contact between rough surfaces]{Wear curves of the contact between rough surfaces. Each curve follows an horizontal line in the wear maps of Figure~\ref{fig:rough_wear_maps}. The filled areas represent the standard deviation. The first non-zero value of each curve is shown by a star. The maximum number of particles reached by each curve is shown by a dot. The invalid parts of the curves (shown crossed in the wear maps) are cut. \textbf{(b)}~The evolution of the wear volume with $\tilde{p}_\text{N}$ notably shows the transition between a regime with zero wear volume (to the left of each star) to a regime where the wear volume increases with the load. All the curves follow the same trend, although with an horizontal shift. For a given curve, between the star and the dot, the wear volume increases steadily. The slope increases drastically around the dot, indicating a transition to a severe wear regime.}
	\label{fig:rough_wear_curves}
\end{figure*}

For an easier interpretation, the wear maps can be represented as curves (Figure~\ref{fig:rough_wear_curves}), where each curve corresponds to a constant value of $d^*$ and vary with the imposed load. Every point of each curve is the average between five measurements done with different randomized rough surfaces, and the standard deviation is indicated. To find the maximum number of particles reached by one curve without being sensitive to the statistical noise, a smoothed version is first computed using a Savitzky-Golay filter of degree 3 on a window of 11 points, and the maximum is determined on the filtered smooth curve. A study of the evolution of the wear volumes (Figure~\ref{fig:rough_wear_curves}(b)) reveals the emergence of three wear regimes:
\begin{itemize}
	\item[--] There is no wear particle production until the normal load reaches a critical value. This range is the `plastic' region: the surfaces are only deformed plastically. This would correspond to the regime of \emph{low wear}.
	\item[--] Above a critical load, wear increases monotonically. This range goes roughly up to the point of maximum number of particles and would correspond the regime of \emph{mild wear}.
	\item[--] For loads higher than the point of maximum number of particles, the slope of the curves increases quickly. This drastic increase of the wear volume would correspond to the regime of \emph{severe wear}.
\end{itemize}

The ability of our model to predict a regime of severe or catastrophic wear is novel among the similar existing models\cite{frerotMechanisticUnderstandingWear2018,brinkParameterfreeMechanisticModel2020}, as these models limit the formation of each wear particle to occur under a single micro-contact. The maximum instantaneous wear volume is thus limited by the size and number of micro-contacts, whereas our model takes into account elastic interactions and permits the creation of wear particles larger than a single micro-contact. One advantage of the model of Brink \emph{et al}. is that it simulates the sliding process, and reaches a steady state wear rate. This procedure is unfortunately not applicable for our model because of the high computational cost to compute even a single pixel of a wear map. In consequence, our model can only predict an instantaneous wear volume and has no notion of sliding distance. Brink \emph{et al.} showed that simulating the sliding history is key to recover Archard's law, stating that harder materials wear less. The current model gives the opposite trend. A computationally efficient procedure to account both for the sliding history, as in Brink \emph{et al.}, as well elastic interactions for wear particle generation, will be the topic of future work.

It is also worth noting that our model predicts the possibility of forming wear particles smaller than $d^*$\snspace, as shown by Figure~\ref{fig:wear_map_2mc} and Figure~\ref{fig:rough_particles}(a) to (d), which is surprising since $d^*$ is generally thought as the minimum possible wear particle size. This effect can also be related to sliding not being considered in our model. As sliding dissipates energy, it would have an effect on the remaining energy available for the creation of such small wear particles. The possibility of forming wear particles smaller than $d^*$ in the absence of sliding is worthy of investigation either experimentally or numerically using FE or MD simulations. In the case where the possibility of creating wear particles smaller than $d^*$ were to be invalidated, our results would not be significantly impacted. The `plasticity` zone in Figure~\ref{fig:rough_wear_maps} would simply extend on a larger range of normal load, and the three identified wear regimes would remain present.

\subsection{Roughness parameters}

The wear maps and curves of the previous section were computed for a unique set of roughness parameters, namely $H = 0.8$, $q_\text{l} = 8$ and $q_\text{s} = n/8$. A study of their respective effects when being varied is reported in Appendix~\ref{apx:rough_params}.

Overall, while the number of produced particles vary with the roughness parameters, the wear volumes remain relatively unaffected. The three previously identified wear regimes also remain untouched. This invariance with the roughness parameters implies that the wear mechanisms are supposedly not affected by the details of the fractal description of the rough surfaces in contact, and that they can be described solely by $d^*$\snspace, which includes the material parameters $E$, $\nu$, $\sigma_\text{j}$ and $\gamma$, and by $\tilde{p}_\text{N}$, which is linked to the imposed normal load $p_\text{N}$ and the RMS of slopes $\sqrt{\langle|\nabla h|^2\rangle}$ of the rough surface. However, this only takes into account an instantaneous measurement of the total wear volume. Actually, the size of the produced wear particles are dictated by the fractal parameters, and size of the detached particles may dictate how the surface roughness \emph{evolves} over time (effectively changing $\sqrt{\langle|\nabla h|^2\rangle}$, thus $\tilde{p}_\text{N}$), and the wear particles themselves may contribute to the tribological properties of the interface, so that their size would be a matter of importance.

\section{Conclusion}

A model of adhesive wear was developed analytically and implemented numerically. The model takes into account elastic interactions between several nearby micro-contacts and allows for the formation of combined wear particles encompassing multiple micro-contacts. A salient result is that a wear particle is not necessarily formed under a single junction, which challenges the definition of what amounts to a contact junction in the context of adhesive wear.  The model is based on two criteria: an energy balance and a crack initiation criteria. It predicts the transition between a regime of low wear (with zero wear volume) to a regime of mild wear, and finally to a regime of severe wear, emerging thanks to the consideration of the elastic interactions. The instantaneous wear volume is predicted from the material parameters, the loading conditions and roughness parameters. Hard materials favor elastic interactions and combined wear particles.

\printbibliography
\balance
\clearpage
\onecolumn
\appendix
\renewcommand\thefigure{\thesection.\arabic{figure}}
\renewcommand\thetable{\thesection.\arabic{table}}
\setcounter{figure}{0}
\setcounter{table}{0}

\section{Appendix}

\subsection{Turning convolutions into cross-correlations}\label{apx:conv_corr}

Following the definition of $u_x$ caused by $p_x$ in the text, we can find any component $i$ of the displacement as
\begin{align}
	u_i(x, y) &= \iint u_{j \rightarrow i}^\text{ker}(x - \xi, y - \eta) \, p_j(\xi, \eta) \,d\xi\,d\eta \\
			  &= [u_{j \rightarrow i}^\text{ker} * p_j](x, y) \,,
\end{align}
which is a convolution, where $i$ and $j$ can be either of the three coordinates $x$, $y$ or $z$ and the Einstein summation convention is used. The full expression of the elastic energy can be written as
\begin{equation}
	E_\text{el} = \frac{1}{2} \int_\Gamma [u_{j \rightarrow i}^\text{ker} * p_j] \, p_i \, d\Gamma \,,
\end{equation}
now taking into account the components of $\bm{u}$ and $\bm{p}$ in all directions.

Alternatively, the integrand $[u_{j \rightarrow i}^\text{ker} * p_j] \, p_i$ of the elastic energy can be written as
\begin{align}
	[u_{j \rightarrow i}^\text{ker} * p_j] \, p_i &= \iint u_{j \rightarrow i}^\text{ker}(x - \xi, y - \eta) \, p_j(\xi, \eta) \, p_i(x, y) \,d\xi\,d\eta \\
	&= \iint u_{j \rightarrow i}^\text{ker}(\xi', \eta') \, p_j(x - \xi', y - \eta') \, p_i(x, y) \,d\xi'd\eta' \,,
\end{align}
which, injected into the $E_\text{el}$ expression, gives
\begin{align}
	E_\text{el} &= \frac{1}{2} \iint [u_{j \rightarrow i}^\text{ker} * p_j] \, p_i \,dx\,dy \\
	&= \frac{1}{2} \iint\hspace{-0.5em}\iint u_{j \rightarrow i}^\text{ker}(\xi', \eta') \, p_j(x - \xi', y - \eta') \, p_i(x, y) \,dx\,dy\,d\xi'd\eta' \\
	&= \frac{1}{2} \iint u_{j \rightarrow i}^\text{ker}(\xi', \eta') [p_j \star p_i](\xi', \eta') \,d\xi'd\eta'
\end{align}
which now contains a cross-correlation, denoted by the $\star$ symbol. Using a lighter notation:
\begin{equation}\label{eq:Eel_corr}
	E_\text{el} = \frac{1}{2} \int_\Gamma u_{j \rightarrow i}^\text{ker} [p_j \star p_i] \, d\Gamma \,.
\end{equation}

\subsection{Effect of the normal load on the calculation of the gain of elastic energy}\label{apx:normal}

Let us consider a surface with micro-contacts, loaded tangentially and vertically. The traction field is
\begin{equation}
	\bm{p} = p_x\bm{e}_x + p_z\bm{e}_z \,,
\end{equation}
and the elastic energy, obtained with \eqref{eq:Eel_corr}, is therefore
\begin{equation}
	E_\text{el} = \frac{1}{2} \int_\Gamma \left( u_{x \rightarrow x}^\text{ker} [p_x \star p_x] + u_{x \rightarrow z}^\text{ker} [p_x \star p_z] + u_{z \rightarrow x}^\text{ker} [p_z \star p_x] + u_{z \rightarrow z}^\text{ker} [p_z \star p_z] \right) \, d\Gamma \,.
\end{equation}
When the micro-contacts are unloaded, they can no longer carry the tangential load, so $p_x$ goes to $0$. However, the normal load remains, so that the unloaded elastic energy is
\begin{equation}
	\Delta E_\text{el} = \frac{1}{2} \int_\Gamma \left( u_{x \rightarrow x}^\text{ker} [p_x \star p_x] + u_{x \rightarrow z}^\text{ker} [p_x \star p_z] + u_{z \rightarrow x}^\text{ker} [p_z \star p_x] \right) \, d\Gamma \,,
\end{equation}
where
\begin{equation}
	u_{x \rightarrow z}^\textnormal{ker} = \frac{1}{4\pi G}\left[ (1 - 2\nu) \frac{x}{r^2} \right] = - u_{z \rightarrow x}^\textnormal{ker}
\end{equation}
and $u_{x \rightarrow x}^\textnormal{ker}$ is given by \eqref{eq:ux_ker}. In the particular case where $p_x(x, y) = p_x(-x, -y)$ and $p_z(x, y) = p_z(-x, -y)$, we have $p_x \star p_z = p_z \star p_x$, so that the unloaded elastic energy becomes
\begin{equation}
	\Delta E_\text{el} = \frac{1}{2} \int_\Gamma u_{x \rightarrow x}^\text{ker} \, d\Gamma \,,
\end{equation}
which is independent of $p_z$. Therefore, in this particular case, the unloaded elastic energy does not depend on the normal load, if it is conserved during the unload of the tangential load. The symmetry conditions on $p_x$ and $p_z$ are fulfilled in the simple analytical cases derived in this paper, and they are also satisfied (approximately) in the case of a contact between rough surfaces, which should be statistically similar upon axial symmetry.

\subsection{Calculation of the elastic energy for a single circular micro-contact}\label{apx:Eel1}

Only the component $p_x$ of $\bm{p}$ is non-zero, and we can write $p_x$ as
\begin{equation}
	p_x(x, y) = c_q(x, y) q
\end{equation}
where $q$ is the value of the uniform tangential load and $c_q(x, y)$ is a function describing the shape of the micro-contact, in this case equal to $1$ when $r = \sqrt{x^2 + y^2} < d / 2$ and $0$ otherwise. $p_x \star p_x$ is easier to calculate than $u_{x \rightarrow x}^\text{ker} * p_x$, which means that we can use \eqref{eq:Eel_corr} to calculate the elastic energy. We have
\begin{equation}
	p_x \star p_x = (c_q \star c_q) q^2
\end{equation}
which is an autocorrelation, calculable geometrically. As $c_q$ is a circle of diameter $d/2$, $[c_q \star c_q](x, y)$ is equal to the area of the intersection between two circles of diameter $d/2$ with a distance $r = \sqrt{x^2 + y^2}$ between their centers:
\begin{equation}\label{eq:autocorr1}
	\mathcal{C}(x, y) = [c_q \star c_q](x, y) = \frac{d^2}{2}\cos\left(\frac{r}{d}\right) - \frac{r}{2}\sqrt{d^2 - r^2} \,,
\end{equation}
where we called $\mathcal{C}$ the autocorrelation of $c_q$. In \eqref{eq:Eel_corr}, this autocorrelation multiplies $u_{x \rightarrow x}^\text{ker}$ \eqref{eq:ux_ker}
\begin{equation*}
	u_{x \rightarrow x}^\text{ker} = \frac{1}{4\pi G} \left[ 2(1 - \nu)\frac{1}{r} + 2\nu\frac{x^2}{r^3} \right] \,,
\end{equation*}
which has a $1/r$ component and a $x^2/r^3$ component. Using polar coordinates and with the help of the \emph{Python} symbolic library \emph{Sympy}\cite{meurerSymPySymbolicComputing2017}, we get the integrals of the products with the components:
\begin{align}
	\int_\Gamma \frac{1}{r} \mathcal{C} \, d\Gamma &= \frac{2\pi d^3}{3} \,, \\
	\int_\Gamma \frac{x^2}{r^3} \mathcal{C} \, d\Gamma &= \frac{\pi d^3}{3} \,.
\end{align}
From those, we easily recover the expression of the elastic energy for a single circular micro-contact \eqref{eq:Eel1}:
\begin{align}
	E_{\text{el}, 1} &= \frac{1}{2} \int_\Gamma \frac{1}{4\pi G} \left[ 2(1 - \nu)\frac{1}{r} + 2\nu\frac{x^2}{r^3} \right] \mathcal{C} q^2 \, d\Gamma \\
						   &= \frac{1}{8\pi G} \left[ 2(1 - \nu)\frac{2\pi d^3}{3} + 2\nu\frac{\pi d^3}{3} \right] q^2 \\ 
						   &= \frac{(2 - \nu) d^3q^2}{12G} \,.
\end{align}

\subsection{Calculation of the approximate elastic energy for two circular micro-contacts}\label{apx:Eel2}

In this case, following the notation of Appendix \ref{apx:Eel1}, $c_q$ is made of two circular regions of diameter $d$ with a space $l$ between their centers and having the line connecting their centers making an angle $\theta$ with the $x$ axis. The autocorellation $\mathcal{C}_2$ of $c_q$ in this case can be written as a function of $\mathcal{C}$ \eqref{eq:autocorr1} for a single micro-contact:
\begin{equation}
	\mathcal{C}_2(x, y) = 2\mathcal{C}(x, y) + \mathcal{C}(x - l\cos\theta, y - l\sin\theta) + \mathcal{C}(x + l\cos\theta, y + l\sin\theta) \,,
\end{equation}
which has a centered component $2\mathcal{C}(x, y)$ and two side components. The centered component simply gives a $2E_\text{el,1}$ contribution to the total elastic energy. The integrals of the products of the side components with the terms $1/r$ and $x^2/r^3$ of $u_{x \rightarrow x}^\text{ker}$ have to be approximated by assuming that $x$ and $r$ do not vary much in the region where the side components of $\mathcal{C}_2$ are non-zero. We have :
\begin{align}
	\int_\Gamma \frac{1}{r} \mathcal{C}(x \pm l\cos\theta, y \pm l\sin\theta) \, d\Gamma &\approx \frac{1}{l} \int_\Gamma \mathcal{C}(x \pm l\cos\theta, y \pm l\sin\theta) \, d\Gamma \\
	&= \frac{\pi^2d^4}{16l} \,, \\
	\int_\Gamma \frac{x^2}{r^3} \mathcal{C}(x \pm l\cos\theta, y \pm l\sin\theta) \, d\Gamma &\approx \frac{\cos^2\theta}{l} \int_\Gamma \mathcal{C}(x \pm l\cos\theta, y \pm l\sin\theta) \, d\Gamma \\
	&= \frac{\pi^2d^4\cos^2\theta}{16l} \,.
\end{align}
Using \eqref{eq:Eel_corr}, we finally get:
\begin{align}
	E_{\text{el}, 2} &= \frac{1}{2} \int_\Gamma \frac{1}{4\pi G} \left[ 2(1 - \nu)\frac{1}{r} + 2\nu\frac{x^2}{r^3} \right] \mathcal{C}_2 q^2 \, d\Gamma \\
						   &\approx 2E_\text{el,1} + 2\frac{1}{8\pi G} \left[ 2(1 - \nu)\frac{\pi^2d^4}{16l} + 2\nu\frac{\pi^2d^4\cos^2\theta}{16l} \right] q^2 \\ 
						   &= 2E_\text{el,1} + \frac{1}{32\pi G} \left[ \frac{\pi^2d^4}{l} + \nu\frac{\pi^2d^4(\cos^2\theta - 1)}{l} \right] q^2 \\ 
						   &= \frac{(2 - \nu) d^3 q^2}{6G} + \frac{\pi d^4q^2}{32G} \frac{1 - \nu \sin^2\theta}{l} \,.
\end{align}

\subsection{Effect of the roughness parameters on the wear maps}\label{apx:rough_params}

\begin{figure*}
	\centering
	\subfloat[]{
		\includegraphics{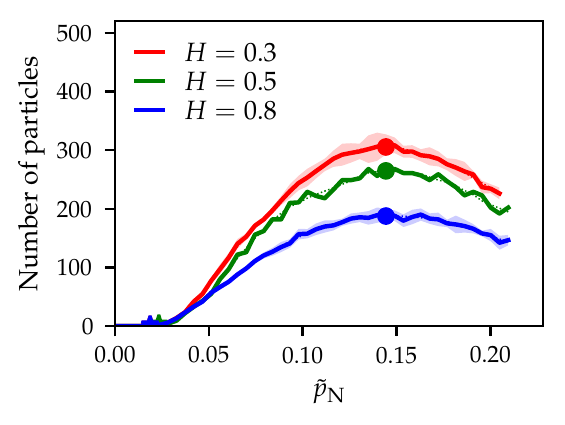}
	}
	\subfloat[]{
		\includegraphics{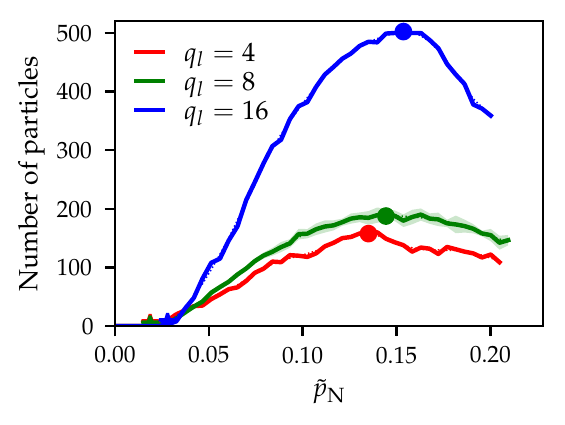}
	}
	\subfloat[]{
		\includegraphics{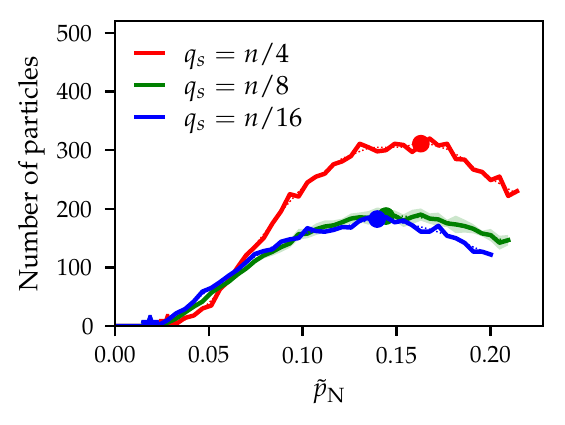}
	}
	\\
	\subfloat[]{
		\includegraphics{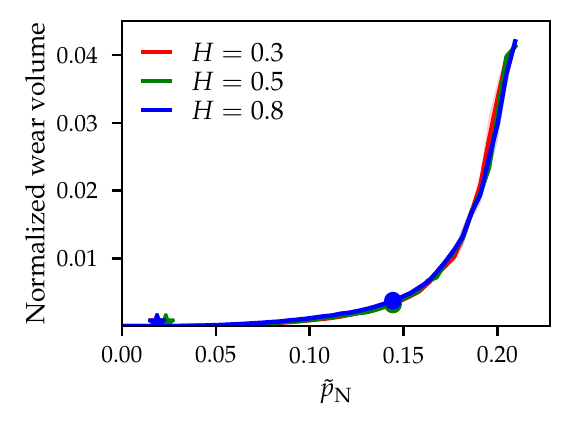}
	}
	\subfloat[]{
		\includegraphics{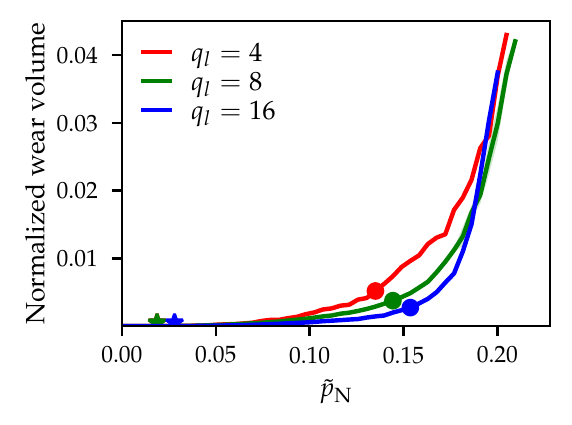}
	}
	\subfloat[]{
		\includegraphics{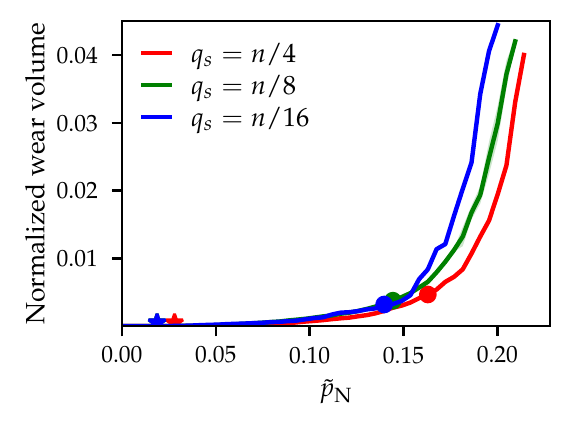}
	}
	\caption[Effect of the roughness parameters on the wear curves]{Effect of the roughness parameters on the wear curves. Here, $d^*/L = 0.07$ and $n = 512$. Only the curves with the sets of parameters $H = 0.8$, $q_\text{l} = 8$, $q_\text{s} = n/8$ and $H = 0.3$, $q_\text{l} = 8$, $q_\text{s} = n/8$ were computed with five repetitions, so their wear curve is averaged and a standard deviation is shown. The curves for the other sets of parameters are computed from only one rough surface and thus are more subject to statistical noise.}
	\label{fig:roughness_params}
\end{figure*}

The effects of the parameters $H$, $q_\text{l}$ and $q_\text{s}$ is assessed by running the simulations listed in Table~\ref{tab:reps} and comparing the curves of number of particles and wear volume. The Figure~\ref{fig:roughness_params} shows such comparison for a single value of $d^*/L = 0.07$ in order to not overload the plots. Note that the RMS of slopes $\sqrt{\langle|\nabla h|^2\rangle}$ is also a roughness parameter, but its effect is already taken into account in the normalized imposed load $\tilde{p}_N$.

\begin{table}[hb]
	\centering
	\caption{List of roughness parameters for the production of wear maps and curves}
	\label{tab:reps}
	\small
\begin{tabular}{cccc}
	\toprule
	$\bm{H}$ & $\bm{q_\text{l}}$ & $\bm{q_\text{s}}$ & \textbf{repetitions} \\
	\midrule
	0.8 &  8 & $n/8$  & 5 \\
	0.5 &  8 & $n/8$  & 1 \\
	0.3 &  8 & $n/8$  & 5 \\
	0.8 &  4 & $n/8$  & 1 \\
	0.8 & 16 & $n/8$  & 1 \\
	0.8 &  8 & $n/4$  & 1 \\
	0.8 &  8 & $n/16$ & 1 \\
	\bottomrule
\end{tabular}
\end{table}

The Figures~\ref{fig:roughness_params}(a) and (d) show the effect of the Hurst exponent $H$. One physical interpretation of $H$ in the context of self-affine rough surfaces is that a surface of size $L$ with a roughness of characteristic height $R$ can be viewed on a window of size $\alpha L$, and the new roughness viewed on this window would have a roughness with characteristic height $\alpha^H R$. It means that for $H = 1$, the surface roughness always look the same in the range of self-affinity (\emph{i.e.} with $q_\text{l} < q < q_\text{s}$) regardless of the scale of observation. A surface with a smaller $H$ will look flatter if zoomed-out and rougher if looked at from a smaller scale. In Figure~\ref{fig:roughness_params}(a), we see that the rough surfaces with a lower $H$ can produce more wear particles, but smaller, as the overall wear volume (Figure~\ref{fig:roughness_params}(d)) is surprisingly independent of $H$. 

The frequency parameters $q_\text{l}$ and $q_\text{s}$ control the region (scaling) of fractal self-affinity of a rough surface. $q_\text{l}$ controls the lower frequencies, so a lower value means higher large scale features. $q_\text{s}$ controls the smaller length scales, so a higher value means smaller rough features. The trends shown by the Figures~\ref{fig:roughness_params}(b) and (c) are in accordance with this description: at higher values of $q_\text{l}$, a rough surface look flatter because the lower frequency shapes are absent, which promotes more contact on the smaller `bumps' on the surface and thus the creation of more wear particles. The trend is the same when $q_\text{s}$ decreases, as more smaller bumps appear and contribute in the rise of the number of wear particles. Still, the total wear volume remains only weakly affected by the change of these roughness parameters.

\end{document}